**Hierarchical network structure as the source of hierarchical dynamics (power law frequency spectra) in living and non-living systems: how state-trait continua (body plans, personalities) emerge from first principles in biophysics**

Goekoop, R. [1]

Kleijn de, R. [2]

NBBR

[1]Corresponding author

Free University Amsterdam

Department of Behavioral and Movement Sciences

Parnassia Academy

Parnassia Group, PsyQ

Department of Anxiety Disorders

Early Detection and Intervention Team (EDIT)

Lijnbaan 4

2512VA The Hague, The Netherlands

R.Goekoop@psyq.nl

[2]Faculty of Social and Behavioral Sciences

Department of Cognitive Psychology

Pieter de la Courtgebouw

 Postbus 9555

 2300 RB Leiden

kleijnrde@fsw.leidenuniv.nl

**Abstract**

Living systems are hierarchical control systems that display a *small world* network structure, in which many smaller clusters are nested within fewer larger ones, producing a fractal-like structure with a 'power-law' cluster size distribution (a mereology). Apart from their structure, the dynamics of living systems also shows fractal-like qualities: the timeseries of inner message passing and overt behavior contain high frequencies or 'states' (treble) that are nested within lower frequencies or 'traits' (bass), producing a power-law frequency spectrum that is known as a 'state-trait continuum' in the behavioral sciences. Here, we argue that the power-law dynamics of living systems results from their power-law network structure: organisms 'vertically encode' the deep spatiotemporal structure of their (anticipated) environments, to the effect that many small clusters near the base of the hierarchy produce high frequency signal changes and fewer larger clusters at its top produce ultra-low frequencies. Such ultra-low frequencies produce physical as well as behavioral traits (i.e. body plans and personalities). Nested-modular structure then causes higher frequencies to be embedded within lower frequencies, producing a power law state-trait continuum. At the heart of such dynamics lies the need for efficient energy dissipation through networks of coupled oscillators, which also governs the dynamics of non-living systems (e.q. earth quakes, stock market fluctuations). Since hierarchical structure produces hierarchical dynamics, the development and collapse of hierarchical structure (e.g. during maturation and disease) should leave specific traces in the dynamics of nested modular systems that may serve as early warning signs to system failure. The applications of this idea range from (bio)physics and phylogenesis to ontogenesis and clinical medicine.

**Keywords**

*Information bottleneck; bowtie; network motifs; morphology; behavior; morphogenesis; body-plans; phenotype; personality; personality development; states; traits; state-trait continuum; pink noise; 1/f noise; Zipf's law; power-law frequency distribution; frequency spectra; Fourier analysis; network systems; hierarchical Bayesian control systems; free energy principle; active inference; network structure; nested-modular; small world; scale free; scale invariant; fractal; mereology; network dynamics; structure-function relationships; nested frequencies; phase-amplitude coupling; living systems; stress; evolution; specialization; speciation; open dissipative systems; thermodynamics; biophysics.*

## 1.0 On the structure (spatial characteristics) of living systems

What causes organisms to have different body plans and personalities? In this paper, we address this question by looking at universal principles that govern the structure and dynamics of living systems. Living systems such as cells, organs, organisms and social networks are known to share a generic network structure that is called a *small world* topology (1). *Small world* networks are a class of networks in which the number of connections per system component (node 'degree') is unevenly distributed across system components, i.e. most nodes have few connections but some have many. For example, most genes, cells, neurons, neural circuits and social individuals have few connections, but some have many (e.g. hub genes, hub neurons, alpha males, community workers). The degree distribution of small world networks follows a characteristic inverse relationship called a 'power-law' degree distribution (2) (Figure 1). This pattern deviates markedly from the Gaussian or 'normal' distributions of attributes such as body weight or height. A well-known hallmark of power-law distributions is that the natural logarithm of the distribution produces a straight line (Figure 1). Another feature is that it is possible to zoom in or out on a power-law curve and still observe the same shape. For this reason, power-law degree distributions are said to have 'scale invariant' or 'scale free' features, which is alternatively referred to as 'self-similarity' or 'fractality' (3, 4).

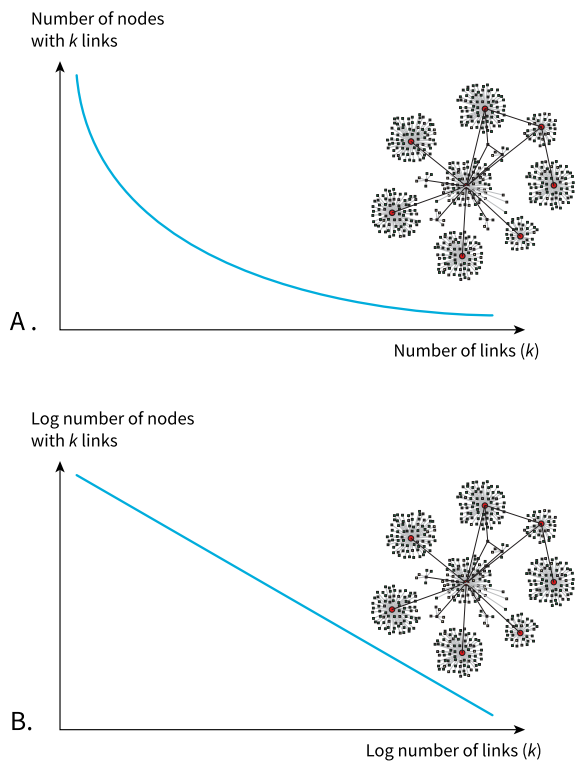

**Figure 1. Living systems: *small world* networks of which the distribution of links across nodes follows a power-law**

*A. The distribution of links across nodes (the degree distribution) in small world networks follows a power-law: most nodes have few connections but some (hubs) have many.*

*B. The natural logarithm (ln) of this power-law degree distribution produces a straight line Y = b\* ln (s), where b is the "power-law exponent", which indicates the steepness of the slope of the line.*

In *small world* networks, highly connected nodes ('hubs') converge onto other hubs to form a hierarchy of hub nodes (a 'rich club' (5)) (Figure 2). This can be compared to teams of horses that are kept in check by a number of horse cart drivers (hubs), which in turn serve as horses that are kept in check by yet higher order drivers (hubs), etcetera, to form a pyramidal structure with a broad base and a narrow apex. As a result of this network topology, messages can travel along hub structures across highly efficient routes, causing any two nodes in the network to be connected via only a small number of intermediate steps (hence the term '*small world*'). Apart from producing efficient pathways, hubs contract parts of the network into so called clusters (modules), which are communities of nodes that share more connections amongst themselves than with their environments (6). This combination of high clustering and low average pathlength is called a *small world* network topology.

In *small world* networks, clusters may themselves serve as hub nodes at a higher spatial scale level of observation that contract collections of other clusters into superclusters and so on, to produce a nested modular, hierarchical structure (Figure 2A). For instance, a set of hub genes in one cluster may connect to hub nodes within other clusters to produce a clustering of clusters. Such nested clustering continues until only a few large modules form the top of a hierarchy of part-whole relationships. A *small world* network topology can be identified in such structures regardless of the spatial scale level of observation, which is why they are called scale invariant or scale free systems (3). When examining the distribution of cluster sizes in such fractal-like systems, a power-law is again obtained (Figure 2B).

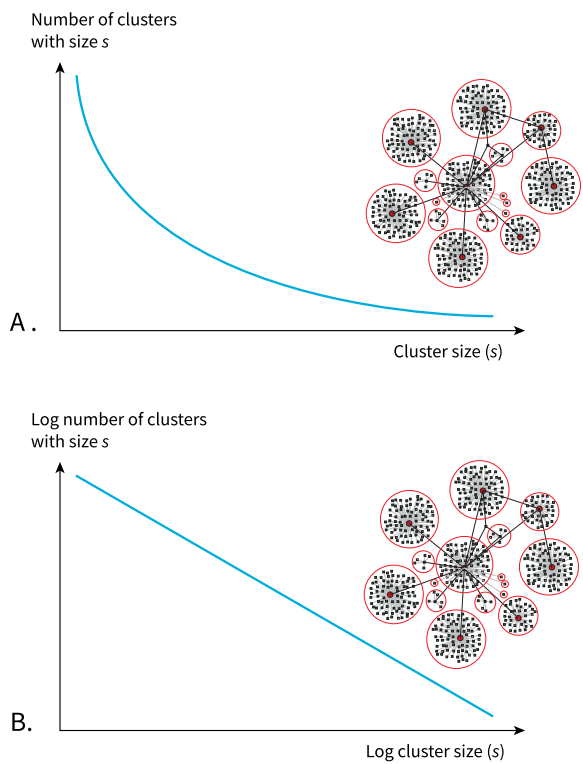

**Figure 2. Living systems: nested modular (hierarchical) network structures of which the cluster size distribution follows a power-law**

*A . In small world network structures, network clusters show conditional dependencies in space, i.e. a given cluster can only exist conditionally upon the existence of its constituent subclusters, producing a spatial hierarchy of part-whole relationships (a 'mereology'). The distribution of cluster size in such scale free networks (e.g. living systems) follows a power-law.*

*B. The natural logarithm of this power-law cluster size distribution produces a straight line.*

**2.0 On the dynamics (temporal characteristics) of living systems**

Apart from their network structure, the dynamics of living systems is known to show signs of scale invariance. The timeseries of inner message passing or overt behavior of living systems shows fast fluctuations that are 'nested' within slower fluctuations, producing a hierarchy of part-whole relationships (7, 8). This can be observed by decomposing the timeseries into their constituent frequencies that are represented by sine waves (this is called Fourier transformation: Figure 3A). Each frequency within this frequency spectrum can be assigned a value that indicates the average amplitude at which that frequency is present within the timeseries. It turns out that the lower (base) frequencies are expressed at the highest amplitudes, after which amplitude smoothly falls off as a power of frequency, producing a typical 'power-law' frequency distribution (Figure 3B, 3C). The various frequency components show conditional dependencies, often such that low-frequency phase changes produce high frequency 'bursts': a phenomenon called phase-amplitude coupling (Figure 3D) (8-10). During task performance, certain intermediate frequencies temporally gain in prominence, which shows up as 'bumps' on the power-law curve (Figure 3E) (7).

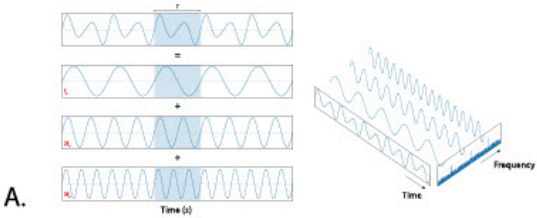

A.

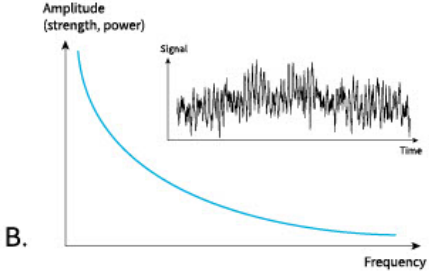

B.

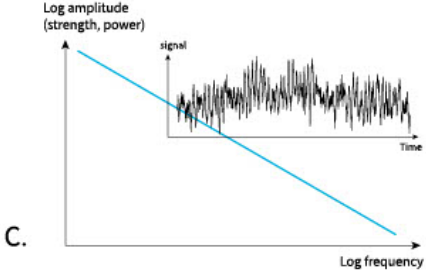

C.

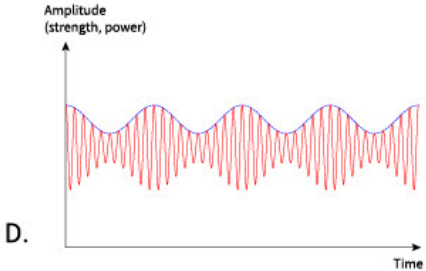

D.

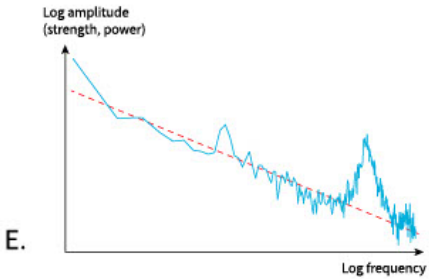

E.

**Figure 3. Living systems show nested modular (hierarchical) dynamics of which the distribution of frequency components follows a power-law**

*A.     Fourier transform is a method by which any signal can be decomposed into its constituent frequencies.*

*B.     The distribution of frequency components of a signal follows a power-law. P ~ 1/ fβ, where P is power or amplitude, f is frequency, and β is a parameter (typically in the range of 0 - 3).*

*C.     The natural logarithm of this power-law frequency distribution follows a straight line.*

*D.     The various frequencies show conditional dependencies. Often, the phase of lower frequencies determines the amplitudes of higher frequencies, so that lower frequencies act as carrier 'waves' for higher frequencies (phase-amplitude coupling or amplitude modulation).*

*E.     Task performance produces 'bumps' on the power-law curve.*

Power-law frequency distributions dominate the timeseries of living systems across a wide range of spatiotemporal scales, varying from sub-millisecond responses of individual photoreceptors and membrane potentials to the distribution of clades across evolutionary timescales (7, 8). They occur anywhere from gene expression and protein synthesis to photosynthesis, respiratory cycles, neural firing rates, neural field potentials and the dynamics of neural systems at large (2). Power-law frequency distributions govern the dynamics of externally observable behavior such as locomotion, posture control, finger tapping, key pressure, reaction time, hit rate, applied force and speech (7, 8). This is also true for the dynamics of subjective inner experience such as mood fluctuations (emotions), motivational changes (mania) and cognitive functions (memory and attention) (11, 12). Power-law dynamics are found in social networks and the internet, making it a truly scale invariant phenomenon (13-15). In essence, a power-law frequency distribution describes a state-trait continuum, in which the rapid fluctuations or 'states' of a system (behavioral 'weather') are superposed onto lower frequencies or 'traits' (behavioral 'climate') (16). A power-law state-trait continuum therefore appears to be a universal feature that governs the dynamics of living systems.

**3.0 On the functional significance of power-law dynamics in living systems**

The ubiquity of the power-law frequency distribution has sparked questions regarding its functional significance. At first, power-laws were considered to reflect measurement artifacts or mere by-products of complex systems, especially since they are equally present in living systems as in abiotic systems such as sand piles (hence the mildly dismissive term 1/f or pink 'noise'). However, recent studies have found evidence to the contrary (7, 8). Changes in scale-free brain activity have been found during development (17, 18), sleep (19), task performance (20) and various physical, neurological and psychiatric disorders (21-24). Such changes involve a flattening or steepening of the power-law curve, indicating a shift in the degree to which different frequencies contribute to the signal at large. Such findings suggest that distinct generative mechanisms underwrite the presence of power-law frequency distributions as well as changes in power-law dynamics during (extreme) task performance, development and disease. Nevertheless, it remains unclear what mechanisms are involved or what degree of universality they might display.

In this paper, we propose that the hierarchical dynamics of living systems (power-law frequency spectra) results from their hierarchical network structure (*small world* network structure). To fully appreciate this relationship, we will first show that living systems can be abstracted as nested modular (hierarchical) networks of coupled oscillators, with an information bottleneck or 'bowtie' motif that allows them to function as hierarchical (Bayesian) control systems. We then show that oscillations in such systems result from circularly causal relationships between excitatory and inhibitory nodes that underwrite a process of hierarchical message passing and predictive coding. Next, we show that organisms 'vertically encode' the deep spatiotemporal structure of their environments to the effect that lower hierarchical regions encode rapidly changing events and higher hierarchical regions encode the slow dynamics of their surroundings. As a result, hub structures at the top of a (regulatory) hierarchy produce slow oscillations (traits) whereas nodes as the base of a hierarchy display fast fluctuations (states), with each intermediate hierarchical level producing its own characteristic frequency (a state-trait continuum). Cross-frequency coupling therefore reflects the coordination between different levels of a nested modular hierarchy, providing a definition for bottom-up and top-down control. We then argue that the typical runoff of amplitude with frequency in power-law spectra results from the fact that living systems are open dissipative systems (i.e. systems that are open to the exchange of energy or matter with the outside world) that are forced to dissipate energy efficiently across multiple hierarchical levels and corresponding frequency bands. The same mechanism has previously been proposed as an explanation for the power-law dynamics of non-living systems, such as earthquake dynamics or stock market fluctuations. Crucially, this means that a single biophysical principle may explain the dynamics of living as well as non-living systems.

Our theory has several remarkable consequences, which center around the fact that the top of a regulatory hierarchy is responsible for producing the stable behavior or 'traits' of a system (i.e. the offsets in timeseries). In living systems, these include the stable aspects of inner experience and overt behavior (personalities) as well as morphological traits (body plans). Small alterations in key regulatory systems at the top of the control hierarchy may produce large changes at its base (behavior), leading to a spectrum of individual differences in personalities and physical traits. Such differences allow organisms to specialize in different econiches. A subtle 'tweaking' of regulatory areas (e.g. point mutations) may therefore suffice to distribute organisms across widely different (social) econiches and optimize survival rates. This provides a principled account of the specialization and speciation of living systems and puts bowties center stage as 'hotspots of evolution'. On a practical note, alterations in power-law frequency spectra may serve as early warning signs to the collapse of hierarchical control, which is a key aspect of many (medical) disorders and physical systems.

**4.0 Living systems as hierarchical control systems**

In a previous paper (Goekoop and de Kleijn, 2021a), we showed that living systems invariably display a nested modular network structure that allows them to function as hierarchical control systems. In this view, the global structure of organisms resembles that of central heating systems: the input to the system (heat) is encoded by the input part of the system (a heat sensor) and compared to a reference state (the setpoint of a thermostat), after which the difference (the error) is conveyed to the output parts of the system (the radiator) to affect the environment (an increase in environmental temperature). This cycle is repeated until the error is reduced to a minimum. This is nicely illustrated by woodlice, which keep on running around erratically until their surrounding humidity levels reach near 100%, which is why we find these creatures in all sorts of nooks and crannies (25). Rather than directly smelling or seeing a damp corner a few yards away and moving directly towards it in a controlled fashion, woodlice keep their motor systems active and vary its output pseudorandomly until they bump into a set of environmental conditions (an 'econiche') that fits their preset reference values (e.g. ~100% humidity), which is when they finally come to rest. Such behavior protects these creatures from desiccation and predation and keeps them stable and intact (homeostasis, survival). Thus, behavior is in the service of producing novel percepts by altering the environment, the aim of which is to bring the organism closer to its reference state: a phenomenon knowns as 'active sensing' or perceptual control (26). This leaves organisms free to generate any kind of behavior that contributes to achieving a specified reference state (e.g. rolling up, digging in or hiding in tight spaces all help to prevent desiccation and evade predators). Such behavioral flexibility allows them to solve many different and unexpected challenges, which greatly adds to their stability (27).

Living systems encode various aspects of their environments by bringing their nodes into specific states. These may be quite simple features (e.g. a single value for humidity or temperature) or a combination of conditions (e.g. humidity, temperature, acidity, glucose levels, light intensity) that together encode a more complex environment or 'econiche' (e.g. the soil of some tropical forest). Such econiches may include social environments, such as mates, friends or rivals. In modeling their environments, the nested modular (hierarchical) network structure of organisms turns out to be ideally suited to solve an important puzzle that every organism faces, which is known as the binding problem (28). In a capricious world, organisms have to decide whether some set of observations is caused by a set of independent causal factors (e.g. three different rivals) or by a single causal factor (e.g. a single mate). In other words, the co-occurrence of several events may have a different meaning and require different actions depending on whether such events have a common underlying (hidden) cause, or rather several independent (hidden) causes. In case of three distinct rivals, these should be encoded separately (segregated, unbinded), whereas a single mate should be encoded as a single factor (integrated, binded). In hierarchical networks, low-level nodes or clusters are used to encode statistically separable contextual cues (e.g. juicy, sweet, green and round), whereas higher-level hub structures encode the instantaneous co-occurrence of these factors, which corresponds to a more integrated yet parsimonious (abstract) model of the environment ('apple'). Even deeper hierarchical layers encode increasingly abstract aspects of their worlds (e.g. categories such as 'fruit' instead of apples, pears and grapes) (29). Thus, nested modular (hierarchical) control systems serve as natural funnels or 'information bottlenecks' (30) that cause the global input of an organism to be encoded by a minimum number of network nodes. This allows organisms to model their econiches with increasing amounts of contextual integration yet parsimony, i.e., abstraction (31). See Figure 4.

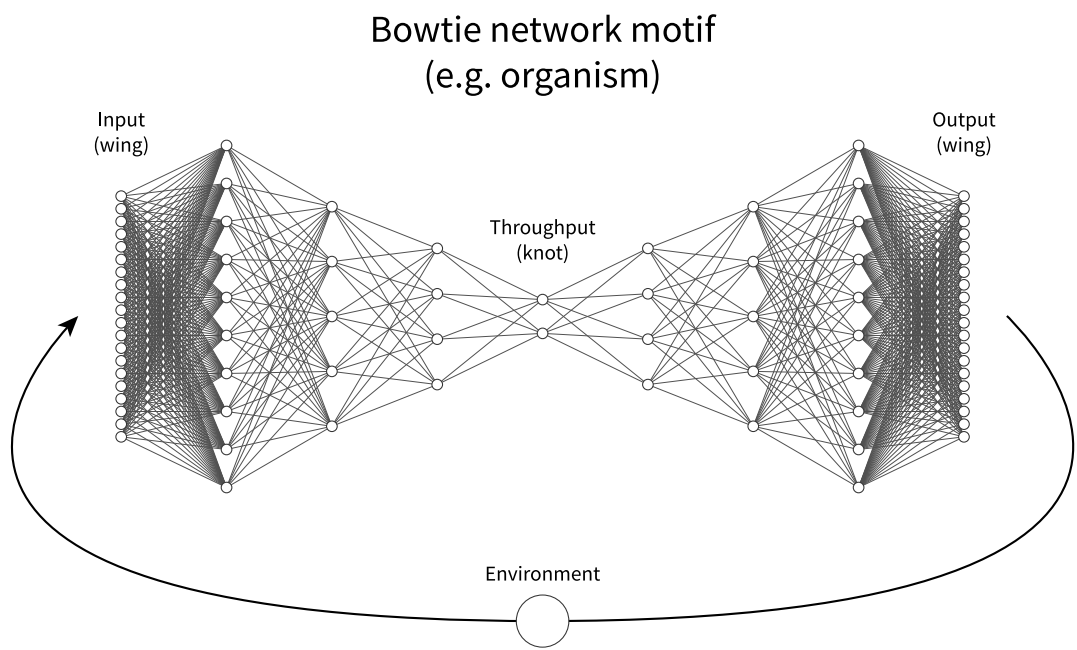

**Figure 4. Living systems as hierarchical control systems**

*Living systems consist of hierarchical (nested modular) network structures with a bowtie-like network motif, in which the top of the hierarchy (the knot of the bowtie) acts as an information bottleneck, producing relatively independent (dual) compartments for input (encoding, perception), throughput (recoding or goal-setting) and output (decoding, action control). The wings of the bowtie (left and right funnels) are involved in perception and action control, respectively. The top of the hierarchy (the knot of the bowtie) compresses information encoded in subordinate layers into long-term and abstract goal states (world models). These models are decoded into sequential output patterns that define the externally observable behavior of the system. The output of the system changes its environment and, hence, the input to the system to form a dynamic loop, in which the bowtie controls the flow of energy between itself and its environment, hence the term 'control system'.*

Organisms do not merely encode the current states of their environments at increasing levels of abstraction, which would amount to perception. Rather, their input layers smoothly transition into throughput layers that encode states (i.e. have values) that deviate increasingly from current reality (Figure 4). Such high level throughput layers (32, 33) encode econiches that are not yet realized but are to be approached (or rather avoided) by generating an output sequence (34). Such (possible) world models or 'goal states' can be compared to the setpoint of a thermostat, which encodes a room temperature that deviates from current reality, making it a model of a *possible* environment (32, 33). Mathematically, the encoding of possible worlds is equivalent to making predictive models of such worlds (33). In this view, the setpoint of a thermostat encodes a prediction of what actual room temperature will be at some point within the future, provided the system will keep on running indefinitely (35). Interestingly, organisms can use their hierarchical structure to make increasingly long-term predictions: apart from using their hub structures to encode the *instantaneous* co-occurrence of separate events as a single higher-level value ('apple'), hubs are used to encode the *sequential* co-occurrence of events (first blossom, then apple). This has the effect of modeling causal relationships (36-38) and anticipating future events (39). When one moves further up the throughput hierarchy, nodes encode increasingly abstract events that are projected ever more distantly into the future (e.g. 'future fruit', e.g. the next harvest). Thus, a 'goal-hierarchy' is produced that constitutes a nested modular causal model with long-term and global goals encoded at the top of the hierarchy, which decompose into a logical set of short-term and specific subgoals (at the base of the hierarchy) that are to be followed in a logical order to reach the global goal (39-42). Thus, when following a goal hierarchy from the top down towards its base near the output hierarchy, global goals are increasingly unpacked (decoded or decompressed) into their constituent subgoals, to eventually form detailed 'output commands' that activate various output organs (e.g. vesicles, flagella, muscles, endocrine glands) (43). This process of unpacking global goal states into executive plans that inform output organs is called 'action control' (40, 42, 44, 45). The subsequent output is a controlled sequence of basic action-perception cycles (behavioral 'primitives' or 'reflexes') that constitutes the overt behavior of an organism. Such behavior serves to alter the environment and, hence, perception, to bring the organism closer to its goals.

In summary, living systems involve nested modular (dual hierarchical) network structures. Such structures naturally form an information bottleneck motif, in which multiple input streams converge onto a few throughput systems (the top of the hierarchy), which in turn diverge onto multiple output systems, producing dedicated compartments for hierarchical perception (input), goal setting (throughput) and action control (output) (46, 47). In systems biology, information bottleneck motifs are called 'bowtie' (2D) or 'hourglass' (3D) motifs because of their physical resemblance to such objects (47). The 'wings' of the bowtie (or the bulbs of the hourglass) produce action-perception pattens (behavior), whereas the 'knot' of the bowtie (or the waist of the hourglass) exerts top-down (goal-directed) control over such patterns. Bowtie motifs engage in a dynamic loop with their environments, allowing organisms to function as hierarchical control systems (Figure 4). The hub nodes that reside at the knot of the bowtie can be compared to generals at the top of a hierarchical chain of command, who acquire a global overview of the battlefield by combining multiple sources of information, after which they need to snap their fingers only occasionally to start a cascade of orders down the executive hierarchy that will eventually make their troops move in different directions (Figure 4). Bowtie motifs have been observed in network systems at all spatial scale levels of observation, including molecular signaling (48), gene regulatory networks (49, 50), neurons, nerves and neural systems (47), whole brains (51), large-scale social networks and the Internet (52). In statistical physics, information bottlenecks are called Markov Blankets. These are limited sets of nodes (hubs) that separate two larger sets of nodes (clusters) into statistically separable compartments (46).

The level of behavioral complexity that bowties can produce depends on their hierarchical breadth as well as their hierarchical depth (39). Broad hierarchies allow for a detailed articulation (factorization, orthogonalization) of context factors. Deep hierarchies allow for high levels of integration across such context factors, producing highly contextualized yet parsimonious ('abstract') models of the world.

Hierarchies that are both broad and deep allow for complex behavioral repertoires that are known as goal-directed behavior (e.g. 'harvesting' requires ploughing the field, seeding the grain, watering the shoots, fertilizing the soil, fending off birds, etcetera, in a logical order). The term 'sophistication' has been reserved for such behavior (39, 53), which includes abilities such as 'agency' (self-functioning, self-directedness, autonomy), social functioning (communion, cooperativeness) and normative functioning (e.g. following moral guidelines or rules) (34, 54). This relates to the cybernetic literature, which speaks of 'homeostatic control' (or system 1) when referring to lower hierarchical levels that control relatively simple processes such as blood pressure or ventilation that are aimed at short-term stability of 'allostatic control' (or system 2) when involving higher hierarchical levels that inspire more complex, effortful and future-oriented behavior that is aimed at securing long-term stability by cycling continuously through different strategies and activities (e.g. collective hunting, foraging, or farming) (36, 39, 55, 56).

**5.0 Living systems as hierarchical *Bayesian* control systems that are engaged in active inference**

The realization that living systems make predictive models of their environments has opened up a whole new field in biophysics known as 'active inference', which aims to produce a first-principle account of the structure and dynamics of living systems (57). According to this theory, the difference between the state of the world as predicted by a reference state and its current value is called a 'prediction error'. As in any control system, prediction error is used to initiate an action sequence that is aimed at altering the environment and (hence) the input to the system, which may reduce prediction error and bring the system closer its goals. In active inference, however, the same prediction error is used to update the predictive model itself (i.e. alter the reference state), allowing the organism to meet environmental conditions halfway. Organisms can therefore reduce prediction error in two fundamental ways: either by changing their environments through action (a process called niche construction (58)) or by changing themselves through model revision (a process called 'belief updating' or 'learning') (59). The combined use of action and model revision allows organisms to iteratively build better models of their worlds (60). This speaks to the active sensing or perceptual control literature (see above), although active inference puts more (cognitivist) emphasis on inference rather than perception per se.

According to information theory, reducing prediction error (improving model fitness) is to reduce a quantity called 'mean variational free energy' (see Box 1). This means that while optimizing their predictive models of the world, organisms are actually trying to descend upon a low-energy stable state (61). By doing so, they follow the second law of thermodynamics, which states that any system that is open to the in- and efflux of matter or energy must seek its lowest possible energy state (i.e. maximum stability) despite a continuous influx of energy or matter. In this respect, organisms are not much different from rivers flowing downstream, ping-pong balls rolling into pits, or coins that start rolling on their sides to reduce friction when dropped to the ground. These systems are all compelled to seek their lowest possible (potential) energy states, exploring a variety of intermediate states or configurations in the process. Such configurations can be seen as 'models' that encode the state of the system's environment (57). Thus, information processing (hierarchical message passing and predictive coding) ultimately involves the dissipation of free energy across time (59, 62, 63), which is as much a physical imperative for living systems as it is for rivers to flow downhill (Box 1). Interestingly, the low-energy stable state that organisms seek in this way is called 'homeostasis' or 'survival' in biology and the process of seeking stability through change is called 'allostasis'. Optimal information processing (building accurate models of the world) is therefore a prerequisite for survival in living systems (64).

**------- Start of Box 1 ---------- Active Inference and the Free Energy Principle**

*According to the second law of thermodynamics, all systems that are open to the in- or efflux or matter or energy must get rid of that energy as efficiently as possible and seek their lowest possible energy*

*states despite a continuous influx of energy or matter (e.g. rivers flowing downhill). In the process, they adopt a spatiotemporal configuration that allows for optimal dissipation of energy, thus forming a 'model' of their environments (e.g. rivers modeling mountains and vice versa) (57, 65, 66). Living systems abide by the second law by trying to reduce their mean variational free energy levels (prediction error), i.e. model error that is iteratively reduced through an active process of perceptual resampling (57). In this process (called active inference), organisms iteratively change their environments (through action) and/or themselves (through model revision or 'learning') to adopt a spatiotemporal configuration (a predictive model) that is optimal in reducing prediction error. The relatively stable state that organisms thus achieve is called 'homeostasis' or 'survival' in biology and the process of seeking stability through change is known as 'allostasis' (67). In statistical physics, low prediction error corresponds to high model fitness, which corresponds to Bayesian (Kalman) filtering or predictive coding (68). Mathematically, active inference can be described as a gradient descent on free energy (61). In information theory, predictive coding or information processing reduces to (metabolic and free) energy dissipation in accordance with the second law of thermodynamics. Interestingly, optimal dissipation of (variational) free energy is achieved by balancing model accuracy (precision) with model complexity (i.e. the number of variables used to explain the data), which yields a model of sufficient accuracy yet parsimony (Ockham's Razor) (63). This 'free energy principle' (FEP) applies to living as well as non-living systems (62, 63, 66). It predicts that hierarchical structures (information bottlenecks, bowties) emerge spontaneously because of a need to optimize (free) energy dissipation, which has indeed been confirmed experimentally (69-73). In this paper, we argue that the same principle also explains the emergence of hierarchical dynamics in living as well as non-living systems, i.e. power-law frequency spectra.*

*The FEP describes a process of filtering out signals that accurately and parsimoniously predict events from signals that carry less reliable predictions (i.e. noise), as a precondition to remaining stable. In doing so, organisms are thought to use a physical implementation of variational Bayesian statistics that allows them to 'reason back' from a sequence of observed events (e.g. some sensory input) to the unobserved (hidden) causes of such events (i.e. the 'latent causal structure' of the observed effects). This process, called 'model inversion', allows organisms to peak through the veil of their imperfect sensory samples at the underlying events that probably caused them. The whole process of estimating the hidden causal structure behind a given input is called Bayesian inference (from Latin 'in-ferre', meaning 'to bring or carry (meaning) into'). As such, active inference can be read as a formal theory of enactivism (74)): a philosophical stance that emphasizes the embeddedness of living systems into a physical (body) and ecological environment (econiche) with which they intimately exchange signals. Such signals are intrinsically meaningless and organisms must actively project meaning into such signals in order to survive, i.e. signals must be identified as reliable predictors of events that may enhance or reduce prediction error (uncertainty), i.e. signals that may ultimately affect their stability and survival.*

**------ End of Box 1 ------**

The equations that describe the process of active inference rely heavily on variational Bayesian statistics, which describe the constant updating of predictive models in the face of novel evidence (e.g. sensory input). At the heart of these equations is the free energy principle, which tells us that the best way to get rid of an excess of (mean variational) free energy is to make accurate yet parsimonious models of the world and vice versa (Box 1). To achieve this, organisms are thought to use a physical instantiation of hierarchical Bayesian inference (75), which is important to the central argument of this paper i.e. that the (power-law) dynamics of living systems is produced by their (power-law) network structure. To substantiate this claim, we will now briefly discuss a network structure that has previously been put forward as a consensus architecture that universally underwrites the process of active inference in living systems.

**6.0 On the structure and dynamics of hierarchical Bayesian control systems**

Figure 5 shows a proposed consensus network architecture that explains the dynamics of living systems in terms of optimized variational free energy dissipation, or active inference. This model combines key findings from graph theory (e.g. (3, 76)), systems biology (41, 47), machine learning (e.g. (77)) and hierarchical predictive coding as proposed by Karl Friston (78, 79). For a more detailed discussion of this structure and its dynamics under stress, see (34, 80). For a mathematical model that approaches this topology, see (81). This model is still under development and may undergo adaptations in the future.

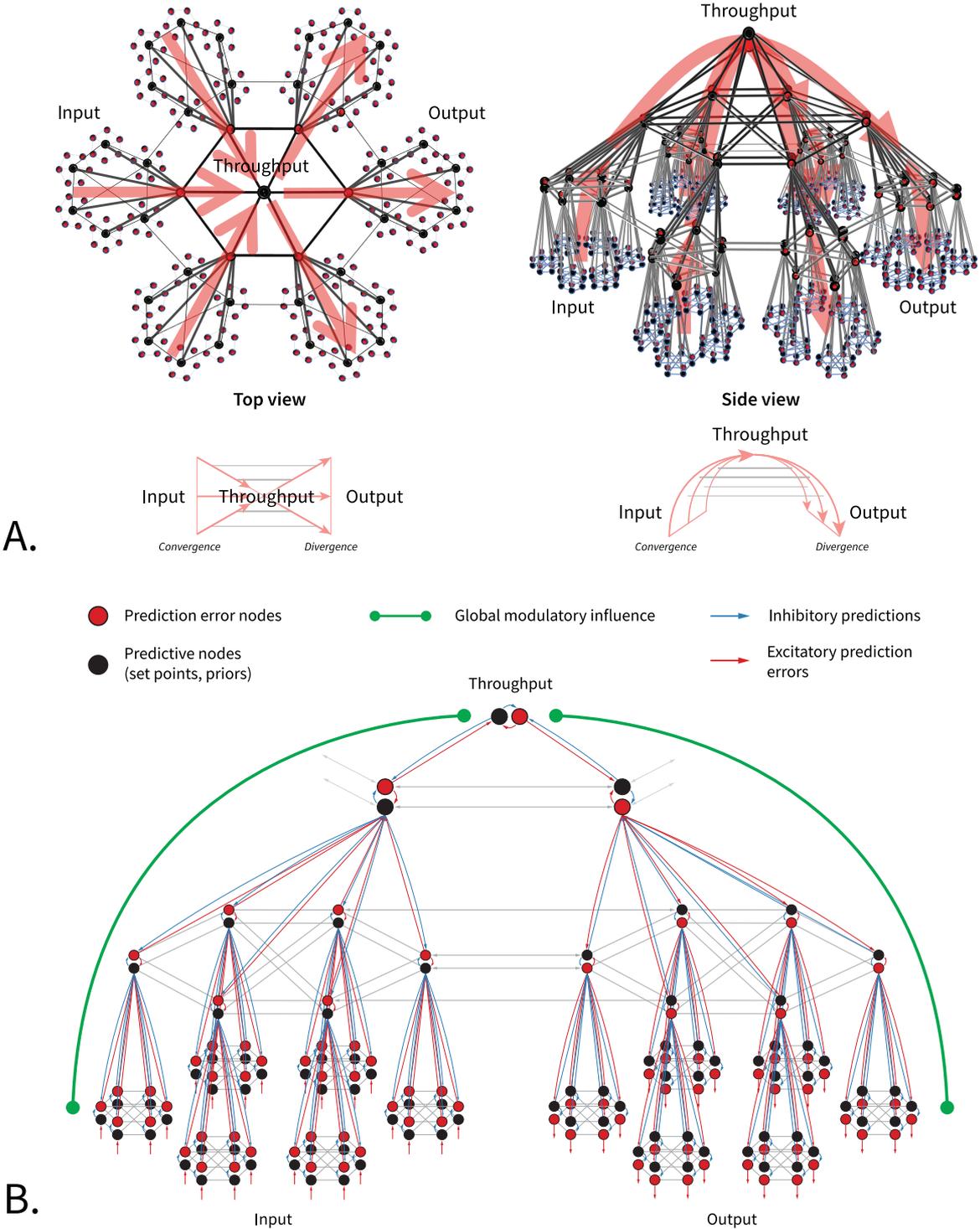

**Figure 5. Schematic image of a network structure that proposes a universal explanation of the dynamics of living systems in terms of optimizing free energy dissipation, or 'active inference'**

*Overview showing a proposed consensus model of a generic network structure that underwrites the process of hierarchical message passing and predictive coding ('active inference') in living systems.*

*A. The global shape of the object is that of a nested modular (dual hierarchical) and folded bowtie network structure. Note the global flow of prediction errors (free energy) from left (input, perception) via the top of the structure (throughput, goal setting) to right (output, executive functioning, and action control). Also, note the global flow of predictions from right (action) via higher level processing (goal setting) to left (perception) to eventually bias perception (selective attention). This global flow is not shown in Figure A for reasons of clarity, but see B. Horizontal connections: skip-connections that bypass higher-level processing and cause the bowtie structure to fold back onto itself. These action-perception loops underwrite automatic responses, the level of complexity of which is a function of hierarchical depth: short cycles that bypass higher level processing correspond to simple reflex arcs, whereas progressively longer loops that visit deeper levels correspond to instinct patterns, habitual behavior, and goal-directed behavior, respectively.*

*B. Excerpt of Figure A, showing nodes and connections in more detail. Black spheres: prediction units (priors, setpoints, predictive models). Red spheres: prediction error units, encoding deviations of some prior value relative to available evidence. Black arrows: inhibitive connections from priors to prediction errors (explaining away evidence). Red arrows: excitatory connections from prediction errors to priors (belief updating). Note the hierarchy of black nodes that encodes a deep (i.e. multilayered) hierarchical predictive model of the inner and outer econiche of the organism (i.e. a world model). Lower levels converge onto higher levels that encode the environment at increasing levels of contextual integration yet parsimony (abstraction). Deeper levels encode progressively slower (stable) aspects of events in the outside world (the vertical encoding of timescale, see text). The hierarchy of red nodes encodes deviations of world models wrt. available evidence (i.e. unexplained evidence). Any unexplained evidence is simultaneously projected upward in the input hierarchy as a residual prediction error that is to be suppressed (inhibited, explained away) by a more sophisticated model of the world (black nodes, see text), as well as relayed to the output (action control) hierarchy to produce an output sequence at a matching level of sophistication. Green bows: global modulatory influences (e.g. neuromodulatory neurotransmitters) that control the overall precision (signal-to-noise ratio, gain) of bottom-up versus top down signaling. See text for further details.*

Figure 5 shows a 'dual' hierarchical (nested-modular) bowtie structure with an input hierarchy (left wing of the bowtie, allowing for perception) that smoothly transitions into a throughput hierarchy (information bottleneck or 'knot' of the bowtie, allowing for high-level goal setting) and an output hierarchy (right wing of the bowtie, allowing for action control). Prediction errors (free energy) grossly flow from input via throughput to output hierarchy to eventually affect output organs, whereas predictions 'flow' in the opposite direction to bias perception. When following the base of the input-hierarchy to its top, network connections converge onto higher-level hub structures that integrate across multiple input streams (functional integration). This forces higher level structures to more parsimoniously encode multiple low-level events (e.g. round, green, sweet and juicy) as a single abstract event (e.g. 'apple'). As observed in the previous section, successive hierarchical levels encode aspects of the environment that are progressively distant from current reality, with the top of the regulatory hierarchy (the knot of the bowtie) encoding the most sophisticated (contextualized, long-term) goal states of the organism. The reverse happens when moving from the knot of the bowtie towards the base of the output hierarchy. Here, connections diverge from high-level hub nodes across subordinate nodes, allowing for a 'decompression' of long-term and abstract goal states into a hierarchical succession of goals and corresponding subgoals to eventually produce detailed 'output commands' that connect to output organs to produce behavior (40, 43). The precision (or positive predictive value) of top-down versus bottom-up signalling can be tuned by global modulatory influences (see below) (82, 83). The bowtie is folded because of 'skip connections' that run horizontally between same-level nodes within the input and output hierarchy, creating shortcuts. These ensure that input signals can skip higher level processing to produce automatic and faster, more energy efficient (but less well-informed) responses. The level of complexity of such action-perception cycles is a function of hierarchical depth: low-level shortcuts correspond to simple reflex arcs whereas progressively deeper loops correspond to instinct patterns (e.g. fight-flight responses), habits and goal-directed behavior, respectively (40).

In Figure 5, a hierarchy of priors (black nodes, setpoints) can be observed that encodes a hierarchical predictive model of the current world that transitions into a goal hierarchy. A hierarchy of prediction error units (red nodes) can be observed next to it that encodes the deviation of predictive models with respect to the available evidence. Prediction errors keep ascending in the input hierarchy until they are sufficiently suppressed (explained away) by a set of higher-level priors. Thus, perceptive models automatically obtain a hierarchical depth and corresponding breadth (a level of sophistication) that optimally matches the available evidence from the outside world. Any residual prediction error then crosses over to the output hierarchy via horizontal skip-connections to activate a cascade of events further down the output hierarchy ('action control') that results in an action sequence (behavior) that may bring the organisms closer to its goals. When sensory input becomes more complex, more prediction error is produced that engages ever higher levels of the goal hierarchy to initiate ever more complex (and long-term) executive plans. As a result, the level of sophistication at which action plans are initiated in the output hierarchy automatically matches that of the perceptual model and corresponding goal state (40). This correspondence of sophistication between perception, goal setting and action control ensures that organism show behavior that matches actual environmental demands, i.e. that organisms produce 'adaptive behavior' (80). This is further exemplified by the fact that predictive loops run backward from action control via goal setting to perceptive hierarchies, representing the top-down biasing of perception (eventually) in the direction of intended action, i.e. 'selective attention' (83, 84). This ensures that organisms pay attention to perceptual cues that are relevant to the planned behavior (85). The selected behavioral policy then changes the environment, which changes the input to the system, after which the cycle repeats. Thus, much like the Baron of Münchhausen who pulled himself out of a swamp by his own hair, living systems save themselves from drowning in a flood of prediction error through a circular process of action and model formation (active inference). Macroscopically, this corresponds to organisms exploring and exploiting their optimal ecological niches (54, 58).

**5.0 Explaining power-law frequency distributions in the timeseries of living systems**

To explain power-law frequency distributions (i.e. system dynamics) from the workings of hierarchical Bayesian control systems as shown in Figure 5 (i.e. system structure), at least four things should be addressed: first, we must explain the process that produces oscillations at different frequencies. Next, we must explain the conditional dependencies that exist between these frequencies as observed in nature, i.e. nested-modular frequency distributions and cross-frequency coupling. Thirdly, we have to explain the inverse relationship between frequency and amplitude in power-law frequency distributions (i.e. why no specific intermediate frequencies dominate the power-law during the resting state). Finally, we must explain the flattening of the slope of the power-law curve ('bumps') during task performance and high levels of stress. These points will be addressed below.

**5.1 Explaining the origin of oscillations at different frequencies in the timeseries of living systems**

In the active inference literature, oscillations arise as a result of circularly causal relationships between priors (setpoints, models) and prediction error units (evidence), where priors inhibit prediction errors ('explain away evidence') and prediction errors excite priors ('perform belief updating') as shown in Figure 5 (78, 79). Although knowledge on oscillatory dynamics in living systems derives from neurodynamics, similar oscillations have been observed in non-neural network systems at various scale levels of observation (62), such as genomes, proteomes and metabolomes (86), calcium waves in networks of pancreatic cells and islets of Langerhans (87, 88), social networks and the internet (89). Through the years, many explanations have been proposed for the various frequencies emergence of such oscillations and various frequencies they display (86). More recently, evidence converges on a key role for hierarchical structure. This will be explained below.

From Figure 5, it can be observed that hierarchical message passing in Bayesian control systems involves prediction errors rather than raw input signals themselves. Each subsequent hierarchical level therefore encodes deviations relative to expected deviations (i.e. prediction errors relative to predicted prediction errors). Mathematically, deviations that change in time can be expressed as derivatives ($dE/dt$), which indicate the amount of change of a certain variable across a small time-interval. When derivatives are hierarchically organized with respect to each other, each derivative encodes a different aspect of the environment. For instance: a change in location indicates speed and a change in speed indicates acceleration. This hierarchical stacking of derivatives produces a vertical distribution in which lower hierarchical levels encode more frequent or 'fast' events (such as a change in location) and higher levels less frequent or 'slower' events (e.g. a change in speed). This is called 'the vertical encoding of timescale' (Figure 6). As a further illustration of this principle, let's consider a simple central heating system in which current room temperature is encoded by a temperature sensor (e.g. 18 degrees) and desired or predicted room temperature by a thermostat (e.g. 22 degrees). Since it may take some time for hot air to travel from the radiator to the temperature sensor (i.e. a conduction delay), the radiator will heat the room beyond a preset temperature before shutting down, causing room temperature to overshoot. Similarly, cold air from the windows may take a while to reach the temperature sensor, producing an undershoot in room temperature before the system kicks back into action. The timeseries of room temperature will therefore show a dampening oscillatory pattern towards the preset temperature (cycles). This may be an unwanted effect, which is why most thermostats contain a so-called 'cycle rate controller'. This is essentially a second setpoint positioned on top of the first: whereas the first (lower-level) setpoint still encodes the user-specified temperature of 22 degrees, the second (higher level) setpoint encodes a reference value for e.g. 50% of the error produced by the setpoint below it. This causes the radiator to shut down early (e.g. when 50% of a positive error has been reduced in case of rising temperatures), or to kick in early (e.g. when 50% of a negative error has been reduced in case of dropping temperatures). A simple stacking of setpoints therefore has the effect of *anticipating* overshoots and undershoots, allowing oscillations to be minimized and room temperature to reach a desired value in a smoother way. In short, higher hierarchical levels predict 'changes in changes', which tend to involve increasingly slow (rare) events in a spatiotemporally structured environment (Figure 6).

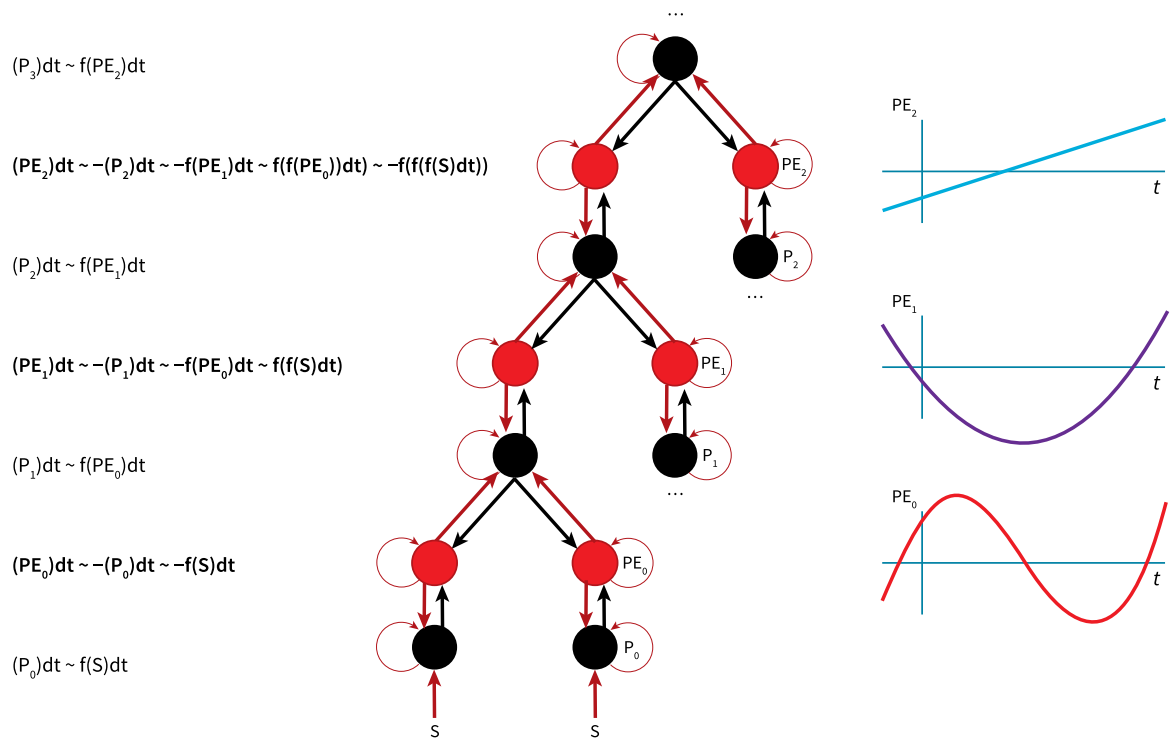

**Figure 6. The vertical encoding of timescale: how frequencies change as a function of hierarchical depth in hierarchical (Bayesian) control systems**

*Excerpt of Figure 5 showing the base of the input hierarchy. P: prior node, PE: prediction error node. S: sensory input. Formulas show how prediction errors at each level globally depend on prediction errors at a lower level (rather than on sensory input per se). As a result, each PE node encodes derivatives of (functions of) derivatives, producing slower dynamics at higher hierarchical levels (i.e. frequency $PE_0 > PE_1 > PE_0$). Note that this is a form of low-pass (Kalman) filtering (90). Thus, a vertically organized gradient of frequencies emerges that encodes the deep spatiotemporal structure of the environment (see text).*

In engineering and cybernetics, the vertical encoding of timescale relates to the good regulator theorem (or the law of requisite variety), which states that to control certain aspects of the environment, a system must encode models that are at least as sophisticated as the aspects of the environment it tries to control (91). In other words, in order to be an effective controller, a control system must encode the deep spatiotemporal structure of its environment and act upon such models. Thus, hierarchical structure allows for a progressive finetuning of a system's behavior by anticipating increasingly rare and abstract events and seek stability (survival) through constant pre-emptive action (i.e. allostasis). For living systems, this means that hierarchical structure conveys an important reproductive advantage. By now, many studies have confirmed that timescale is encoded hierarchically as gradients across the brains of many different species, including nematodes rodents and primates, revealing it as a general organizing principle of brain function (92-98). In humans, mesocortical structures form an information bottleneck that encodes slower dynamics than its wings (the sensorimotor cortices) (99). A hierarchical encoding of timescale has further been observed in molecular networks (100) decision making processes in human organizations (101), suggesting it is a scale free phenomenon.

Due to the hierarchical encoding of spatiotemporal structure, each level in a nested modular network structure has its own characteristic frequency. In most natural systems, hierarchical levels are not as neatly stratified as shown in Figure 5, in which each scale level can be assigned a discrete number (scalar dimensions). Rather, hierarchical scale levels smoothly dissolve into each other, causing a continuous or 'broken' dimensional scale (hence the term fractal). This may explain why Fourier analyses tend to produce a smear of frequencies (i.e. frequency *spectrum*) rather than a set of discrete spikes. In the next paragraph, we will examine the nature of the relationships between the various frequencies that are produced in hierarchical (Bayesian) control systems.

Interestingly, simulation studies show that non-randomly connected (nested modular, *small world,* hierarchical) network systems spontaneously engage in a vertical encoding of timescale from the sole requirement of optimized energy dissipation (102). Since hierarchical structure itself may develop spontaneously as a result of the optimization of energy dissipation (46, 69-73), this suggests that optimal energy dissipation may be the sole requirement for the emergence of hierarchical network structure as well as dynamics (i.e. power-law frequency spectra). This argument will be further pursued in paragraph 5.3.

**5.2 Explaining nested modular frequency distributions (with cross-frequency coupling) in the timeseries of living systems**

Experiments show that the frequency components of the timeseries of living systems are far from independent. A burgeoning neuroimaging literature highlights the importance of cross-frequency coupling in coordinating brain activity across spatial and temporal scales (103). In many cases, the phase of the lower frequencies controls the amplitude of the higher frequency components in the human brain (104, 105) (Figure 2). Thus, low frequency oscillations may initiate high-frequency 'bursts' and vice versa. In our view, cross-frequency coupling can be explained by the mereological couplings (spatial dependencies, part-whole relationships) between the nodes and clusters of nested modular network systems: since higher level clusters can only exist conditionally upon the presence of their constituent subclusters, their activity must be conditionally dependent as well, producing nested frequencies (Figure 7). Simulations studies show that such mereological dependencies indeed produce nested modular (power-law) dynamics (106), even in the absence of criticality (81). The exact mechanism by which phase-amplitude coupling occurs remains unclear, however, although our theory agrees most with a previously posited explanation stating that lower frequencies modulate the gain (precision) of circularly causal loops between excitatory and inhibitory units (103, 104, 107) in ways comparable to selective attention (108). Simulations show that this causes lower frequencies to shape the trajectories of higher frequencies, to automatically produce a phase-locking between slower and higher frequencies (103, 104, 107).

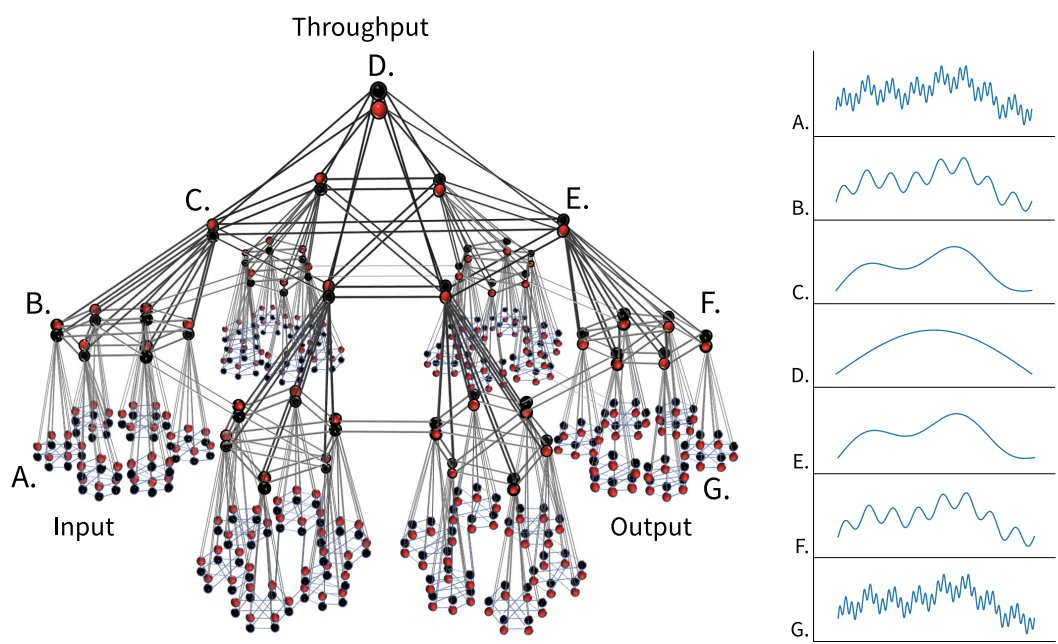

**Figure 7 How hierarchical structure produces hierarchical dynamics: power-law frequency spectra reflect hierarchical message passing in nested modular systems**

*In nested modular systems, each hierarchical level produces its own characteristic frequency (see figure 6). Cross-frequency coupling therefore represents the communication between hierarchical levels (top-down versus bottom-up control). Because of their mereological dependencies (part-whole relationships), amplitude changes at level A (the signal) depend on amplitude and phase changes at level B (the carrier wave) and so on. Thus, higher frequencies become nested within lower frequencies, explaining the nested modularity of the frequency components of powerlaw frequency spectra. Only cross-frequency dependencies of amplitudes are shown. Cross-frequency phase-amplitude modulation is omitted in this figure for visualization purposes. See text for further details.*

In our view, each level within a nested modular hierarchy has its own characteristic frequency, with lower frequencies being produced at higher levels and higher frequencies at lower levels. Cross-frequency coupling should therefore reflect the coupling between different levels in a hierarchical control system. This can be read as a definition of top-down and bottom-up control (109): like generals forming a grand overview of the battlefield based on which they issue out highly global orders, hub nodes and clusters (at the top of a bowtie) encode global goal states that manifest as low-frequency signals. Such signals subsequently control lower-level signals through a cascade of interlocked frequencies until they shape the trajectories of action-perception cycles at the base of the hierarchy (the wings of the bowtie), defining the process of top-down control (e.g. action control or perception control / selective attention). Higher hierarchical levels essentially impose their tonicity onto the output of their subordinate systems, so that the timeseries of low-level action-perception cycles (overt behavior) incorporate the frequencies of all superordinate levels that were involved in controlling the output of the organism. This would explain the universal presence of power-law dynamics in the behavioral timeseries of living systems, as discussed above (110). A similar process may define bottom-up control, in which high frequency signals at lower levels shape the trajectories of low-frequency signals within higher regions, reflecting the process of belief updating in the face of novel evidence (i.e. novel percepts). Cross-frequency coupling should therefore have different (causal) directions depending on whether one examines the input/perception, output/action or throughout/goal-setting parts of a control system. This is in line with results from numerical as well as simulation studies based on empirical data, showing that high-degree (high-level) hub nodes in biologically plausible networks may both phase-lag and phase-lead signal changes in lower degree nodes (111). A long history of psychometrics further shows that power-laws-govern human inner experience just as well as overt behavior: state-trait continua have been observed in the timeseries of nearly every aspect of the human mental phenotype, varying from attention levels global experiences such as self-image or theories of mind, e.g. (112, 113). We expect that the hierarchical network structure of the human brain explains the state-trait continua of subjective experience in the same way as they do for overt behavior (Figures 6 and 7 and see below).

In a seminal paper on the subject, He et al consider the possibility that the power-law (broadband) signal results from the summation of many narrowband oscillations, precisely as we propose, but dismiss it as 'near magic', stating that 'overwhelming evidence' now suggests that there are at least two distinct phenomena in the brain, which are the 1/f power-law activity and the rhythmic oscillations often studied using EEG or MEG recordings which show up as bumps on the power-law curve (7). We believe that this view rests upon a notion of the brain prior to the full recognition of its nested modular topology and precise knowledge of the predictive coding and active inference framework as discussed above, which ultimately concerns the optimized dissipation of mean variational free energy across the full spectrum of frequency bands. This will be discussed below.

**5.3 Explaining the inverse amplitude-to-frequency relationship in the timeseries of living systems**

As shown above, living systems can be abstracted as networks of coupled oscillators that are in the act of dissipating (metabolic or free) energy, which is what defines information processing. The energy that is required to produce oscillations is supplied by metabolic pathways that ultimately require glucose and ATP (114)). During hierarchical message passing, sensory prediction errors (variational free energy) drive nodes that use metabolic energy to update internal states and drive the output of the system (active inference; see above). Information processing in living systems can thus be understood as a process by which metabolic energy dissipation (oscillations) is put in the service of variational free energy dissipation (prediction error reduction). In non-living systems, the metabolic component is lacking and nodes simply pass on the energy from their surroundings to other nodes within their network structure until it leaves the system.

Empirical studies as well as computer simulations show that highly connected (hub) nodes oscillate at higher amplitudes (115), although some studies find more variability (111). This observation fits well with our prediction that hierarchical (nested modular) network structure underwrites the power-law dynamics of living systems: in *small-world* networks, the most strongly connected nodes (hubs) reside at the top of the hierarchy (the knot of the bowtie), where they emit strong oscillations at low frequencies. When moving down the hierarchy, coupling strength (weighted node degree) diminishes in a power-law fashion while amplitudes fall and frequencies rise, which should produce an inverse amplitude-to-frequency relationship. In other words, the non-egalitarian coupling of nodes and modules in *small world* networks imposes a non-randomness of amplitudes onto their frequency spectra, with lower frequencies (produced by hub units) dominating in amplitude over higher frequencies (peripheral nodes). We believe this to be the first time that the universal runoff of amplitude with frequency in power-law frequency spectra is explained in terms of the non-randomness of coupling strength of oscillators in nested modular (scale free) networks and, hence, hierarchical network structure (see Figure 1).

The relationship between coupling strength and oscillation amplitude has been largely observed in simulation studies and is (partly) backed up by empirical studies (111, 115). Nevertheless, it currently lacks a sufficient explanation. In our view, the smooth run-off of amplitude with increasing frequency can be explained from the perspective of energy dissipation in networks of coupled oscillators (see Box 1). In such networks, strongly coupled (constrained) oscillators are 'rigid' oscillators, i.e. they have a faster relaxation rates than loosely coupled nodes, which means they show a quick recovery after perturbation (116, 117). Relaxation rate is directly related to the dissipation of (metabolic and/or free) energy (118). Thus, hub units dissipate large amounts of energy, but are relatively few in number. In contrast, peripheral nodes emit less energy but come in larger numbers. Thus, a small number of energy-demanding hubs with fixed low frequencies dissipate the same amount of energy as a larger number of energy-efficient (peripheral) nodes with fixed high frequencies, causing hubs to oscillate at higher amplitudes than peripheral nodes. In other words, 1/f noise can be explained by the necessity of open dissipative systems to distribute energy dissipation equally across its constituent coupled oscillators, hierarchical levels and corresponding frequency bands.

This proposal is in line with seminal work done by Per Bak, who showed that the size of cascading events in simulated sandpiles displays a 1/f frequency distribution (119). Whereas Per Bak concluded that a self-organized critical state is responsible for producing power-law scaling and even doubted the generality of his models' applicability, it seems that this model can nonetheless be generalized by explaining such cascades (e.g. avalanches in grains of sand, neural avalanches or snow avalanches) as special cases of granular systems involved in the optimization of (free) energy dissipation. This idea receives support from a recently published universal model of earthquake statistics, which uses a simple mean field model to define the coupling between granula (120). According to this model, the earth's crust can be modelled as a network of coupled oscillators of different sizes, varying from (a few) large chunks of earth to a large number of smaller rocks and grains of sand that use vibration and friction to dissipate kinetic energy in the form of heat. The heavier chunks of rock act as (mass-constrained) 'rigid oscillators' that vibrate at lower frequencies and higher amplitudes than the lighter granules and grains of sand, producing 1/f noise. This model is consistent with neurobiological findings that show that hierarchical structure promotes self-sustained 1/f noise by interconnecting modules that display self-organized criticality (106). In this view, the brain is a nested modular system in which each node or module is like a (metabolically activated) granule of a particular size that can be excited by internal or external energy to produce scale-free (avalanche) dynamics at a global level. Thus, a single (mean field) model can be used to explain the dynamics of living as well as non-living systems. Interestingly, mean field models connect directly (through the Bogoliubov inequality) to the equations of motion that describe the process of active inference (121). It therefore seems that efficient energy dissipation ultimately drives the dynamics of living as well as non-living systems. We believe this is the first time that a single physical explanation is given for power-law (1/f) dynamics of living as well as non-living systems.

Previously, been proposed that (changes in) power-law frequency spectra represent (shifts in) the excitation/inhibition balance in the human brain (122). This is in line with another proposal that power-law frequency distributions in EEG signals result from fluctuations in the precision of prediction error and predictive signals (108). In this view, precision is encoded by the amplitude-to-noise ratio of such signals, which is modulated at the synaptic level by neuromodulatory neurotransmitter systems that affect the gain of the loops that run between excitatory (prediction error) and inhibitory (predictive) units (this is represented by the global arches in Figure 5). In our view, shifts in the global excitation-inhibition balance may certainly explain power-law dynamics, but only to the extent that they alter the balance between top-down (predictive, inhibitory) and bottom up (corrective, excitatory) processing and, therefore, the overall hierarchical depth or level of sophistication of information processing. Thus, to the effect that such shifts alter the overall hierarchical depth or sophistication of neural processing (e.g. through cross-frequency coupling), we agree that this may affect the slopes of power-law curves. However, we consider fluctuations in the excitation/inhibition balance by themselves to be insufficient to explain the full phenomenology of power-law frequency spectra, which in our view requires the idea of hierarchical structure.

In living systems, most metabolic energy is consumed by a small number of hub structures at the top of a regulatory hierarchy (so called rich clubs (5, 123)), which encodes highly compressed world models (i.e. long-term predictive models of abstract events). This means that producing such models is allowed at the expense of a metabolic penalty, putting an upper bound on the level of compression or sophistication an organism can achieve when encoding its environment. The 'costs of compression' can be read from the total amount of spectral energy involved in encoding some aspect of the environment, which is given by area under the power-law curve $E = 1/(B-1)*\rho*y^{-(B-1)} + C$, with B being the power-law exponent and C a constant. Theoretically, the area right of any vertical line (x = f) is a measure of the total amount of spectral energy required to encode some aspect of the world up to the hierarchical level that is given by this index frequency. The area left of this line represents the amount of energy required to construct even more abstract or compressed models. The ratio between these two areas is a measure of the metabolic costs per time unit that an organism is prepared to spend on making more sophisticated models of the world than those given by the index level. The well-known deviations of naturally obtained frequency spectra from perfect power-law curves near their asymptotes (76) can be explained by the fact living systems are not perfectly scale free hierarchies: their scale levels have a lower bound at the level of molecules and an upper bound at the level of organisms or ecosystems. This prevents a scenario where infinite amounts of metabolic energy are required by the apex of a hierarchy to compress infinite amounts of uncompressed information encoded at its base.

In summary, we propose that the steepness of the power-law curve conveys information about the hierarchical depth of the generative models that a system encodes: steep curves signal deeper hierarchies whereas flatter curves indicate flatter hierarchies. This means that power-law frequency spectra should change as a function of the hierarchical depth of a generative model as observed e.g.

during development, task performance, stress, and physical or mental illness. This will be discussed below.

**5.4 Explaining changes in power-law frequency spectra during development, task performance, stress and disease**

In task-negative conditions, intrinsically generated stochastic noise produces prediction errors that are escalated upwards in the hierarchy to be suppressed by predictive models. This occurs along the full depth of the regulatory hierarchy, producing a (day)dreamlike state that reflects the free exploration of the network's state space in the absence of external stimuli (106, 124). In our view, this 'resting state' explains the emergence of the full breadth of the power-law frequency spectrum, as well as the characteristic non-dominance of any intermediate component. During task performance, however, prediction error is projected upwards in the hierarchy to a hierarchical depth and breadth that is optimal in solving that task (see above and Figure 5). Thus, task performance usually recruits an intermediate hierarchical level that lies somewhere in between the base of the hierarchy (minimal effort) and its top (peak performance). Since each hierarchical level has its own characteristic frequency, a switch from a resting state to task performance should therefore increase the amplitudes of the frequency components that correspond to the hierarchical levels that are involved in solving that particular task (this is because external stimuli (task performance) produce stronger oscillations than intrinsic noise (resting state)). This may explain why power-law frequency distributions develop 'bumps' at intermediate frequencies during task performance (e.g. alpha, beta, or gamma bands in EEG (8)). (Figure 3). The amplified frequencies that make up these bumps should reflect the hierarchical depth of the chosen model and, therefore, task difficulty.

In this view, power-law frequency spectra should covary with task difficulty: simple tasks that require tax only the base of a regulatory hierarchy (i.e. reflexes or habits) should produce bumps at higher frequencies, whereas increasingly demanding tasks should increasingly recruit lower frequencies, leading to a more global flattening of the power-law curve (Figure 7). This has been confirmed experimentally by examining the effect of varying task loads on the steepness of power-law curves (125, 126). fMRI studies show that difficult tasks recruit lower frequencies in brain areas that are involved in interregional communication (i.e. high-level hub structures) (126). This also relates to the band-filtered EEG literature on the theta-beta ratio, which shows a shift towards the slower theta component when tasks become more demanding (127, 128). Bacteria respond to varying growth conditions primarily by shifts in low-frequency components within the timeseries of gene expression, which is indicative of high-level regulatory changes (129). An increase in bass frequencies as a function of task difficulty therefore appears to be quite universal. From the perspective of active inference (see above), bumps on a power-law frequency distribution signal barriers to the efficient dissipation of variational free energy that need to be overcome by action (task performance) and belief updating (learning). Once the problem is solved, the bumps are 'smoothed out' and organisms can return to a low-free energy resting state (homeostasis, i.e. a power-law curve without bumps, Figure 3).

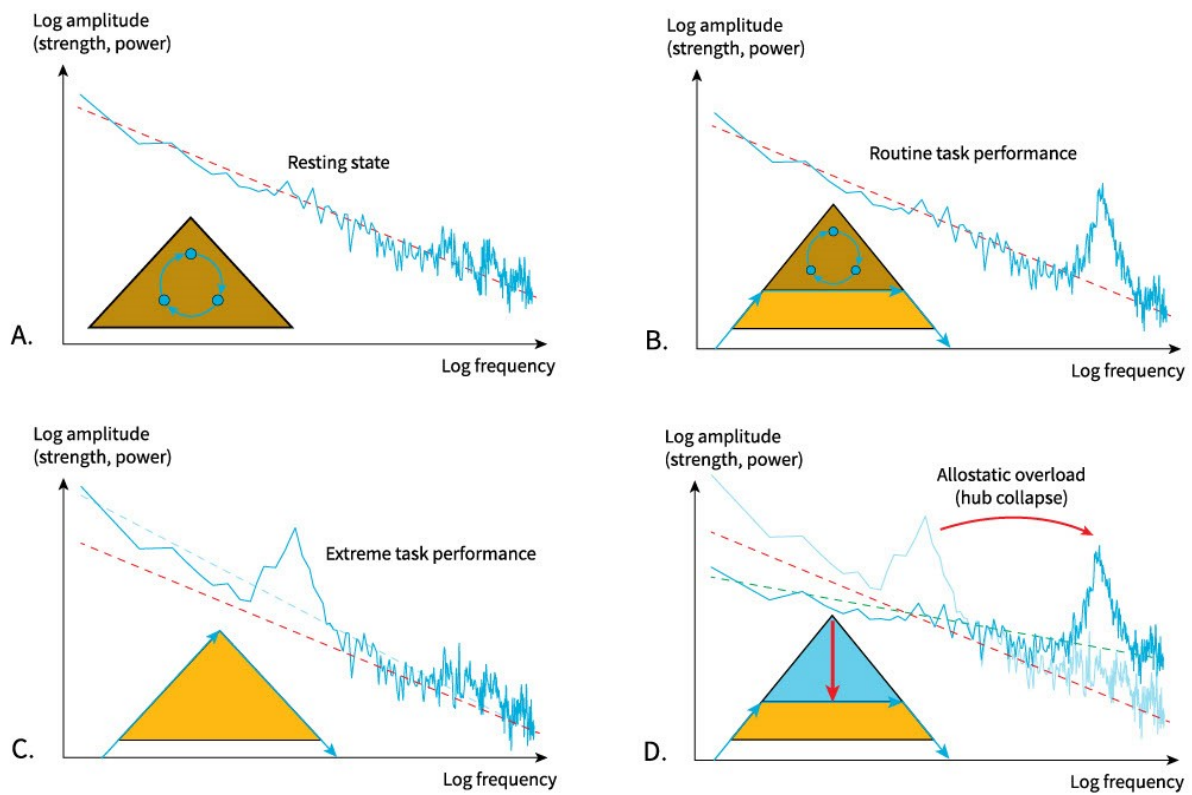

**Figure 8. Changes in power-law frequency distributions as a function of task performance (levels of stress)**

*Increasingly challenging tasks tax increasingly deeper layers of a hierarchical control system until the system collapses in a top-down fashion due to continuous peak performance. We predict that such changes show up as specific alternations in power-law curves. A. Resting state: no bumps are observed. Power-law frequency distribution provides information on the full depth of the structural hierarchy. B. Simple tasks produce high-frequency bumps (flat tail flattening), indicating increased engagement of lower hierarchical levels in (routine, homeostatic) task performance, i.e. reflexes or habits. C. More demanding tasks recruit intermediate tot higher levels of the hierarchy, causing bumps at intermediate to lower frequencies (mid-tail flattening) that reflect more sophisticated (goal-directed, allostatic) forms of control. These frequencies increase in amplitude (reflecting increased recruitment) until D. A tipping point occurs that signals a top-down (metabolic) collapse of the hierarchy as a result of hub overload and cascading failure. This process defines a stress-induced collapse of high-level sophisticated (goal-directed, central integrative, allostatic) control: a phenomenon known as allostatic overload, see text). We predict that this collapse manifests as a discrete event (a tipping point) that coincides with a sudden flattening of the power-law curve (i.e. a drop in low-frequency power, corresponding to a reduced expression of sophisticated traits (e.g. changes in goal-directed personality traits during acute episodes of mental illness), while the system falls back to less sophisticated means of coping with the problem (habitual or routine reflexive behavior), as indicated by an increase in flat tail (high frequency) amplitude. The loss of top-down central integrative synchronization of lower hierarchical levels by higher level hub regions decreases the predictability of activity changes at lower levels. The ensuing unpredictability of the timeseries can quantified in terms of permutation entropy, which is a measure of 'disorder' as unpredictability (see text).*

In a previous study, we proposed that extremely challenging tasks produce prediction errors that reach the top of a regulatory hierarchy, reflecting the peak performance of the system. In this case, the organism must deploy its most sophisticated world models and corresponding action strategies to escape a difficult situation. Continuous peak performance may then cause the regulatory capacity of a system to overload. This is due to the region's disproportionately high density of hub units, which have the highest rates of energy dissipation and, thus, the highest (metabolic) energy demand (see above). The knot of a regulatory hierarchy is therefore most vulnerable to metabolic energy depletion. When energy demand exceeds energy supply, hub units shut down and fail in a cascading manner as a function of node degree, causing a top-down collapse of hierarchical control (34, 80). This coincides with a shift in behavior from so called 'slow' to 'fast' survival strategies, i.e. from high-level (allostatic, goal-directed, integrated, contextualized, abstract, socially inclusive and long-term) to low-level (homeostatic, habitual or reflexive, segregated, decontextualized, concrete, self-centered and short-term) strategies (130). Together, such changes are referred to as 'allostatic overload' in the biological literature (67). Allostatic overload (hub collapse) may continue until a tipping point is reached where large proportions of the network become functionally segregated, marking a discrete transition from a centrally coordinated and ordered state towards an uncoordinated, disordered state that is associated with malfunction, disease, or death (i.e., a tipping point). The ensuing disorder (a loss of predictability) can be quantified by a single term called permutation entropy and turns out to be a hallmark of many physical and mental disorders (80), see below.

Since power-law frequency spectra may convey information about the hierarchical status of a system, the collapse of hierarchical structure or function (allostatic overload) should coincide with specific changes in curve characteristics. This is confirmed by a growing number of studies reporting temporary or permanent changes in power-law frequency distributions in various physical (131, 132) and mental disorders (133-136) as well as during social isolation (137), various sleep stages (138), development (139, 140) and aging (141, 142). The combined results of these studies appear to support the conclusion that increased task difficulty (such as REM sleep versus deep sleep or social engagement versus social isolation) involves a steepening of power-law curves, i.e. a gain in the amplitude of lower frequencies. In our view, this indicates the (increased) recruitment of deeper hierarchical layers that allow for more sophisticated (goal directed, allostatic) forms of control. Conversely, less difficult tasks appear to involve a flattening of the power-law curve (a relative decrease in lower frequency amplitudes), which is consistent with a greater reliance on lower hierarchical levels of control (e.g. habits, instinct patterns or reflexes). For instance, a loss of high-level regulatory hubs (allostatic control) has been observed in diabetes mellitus (143), neurological disorders (144), major depression or schizophrenia (145). This coincides with a flattening of power-law curves in such disorders (133-136). Also, flatter power-law curves have been found in disorders in which regulatory hierarchies fail to reach sufficient depth during development, producing a permanently underregulated state (such as autism spectrum disorders, personality disorders or ADHD) (135, 146). In some cases, however, pathology is related to a steepening of the power-law curve, e.g. during traumatic brain injury or aging. This may reflect situations in which higher (prefrontal) level are recruited (effortful processing) to compensate for the loss of intermediate-level functions. Steeper slopes may occur also occur in cases where aging or disease interfere with the necessity to keep certain frequencies tightly within limits, e.g. heart rate variability (131). The increased involvement of lower frequencies may signal a regulatory failure of systems that keep such variability under control. Overall, four logical types of regulatory deficits may occur in any type of organism across its lifespan, each of which may have both internal and external causes (i.e. control may fail in the presence or absence of extreme environmental circumstances). Such deficits should translate into specific changes in power-law frequency spectra that can be systematically parameterized using existing techniques (147) (Figure 9).

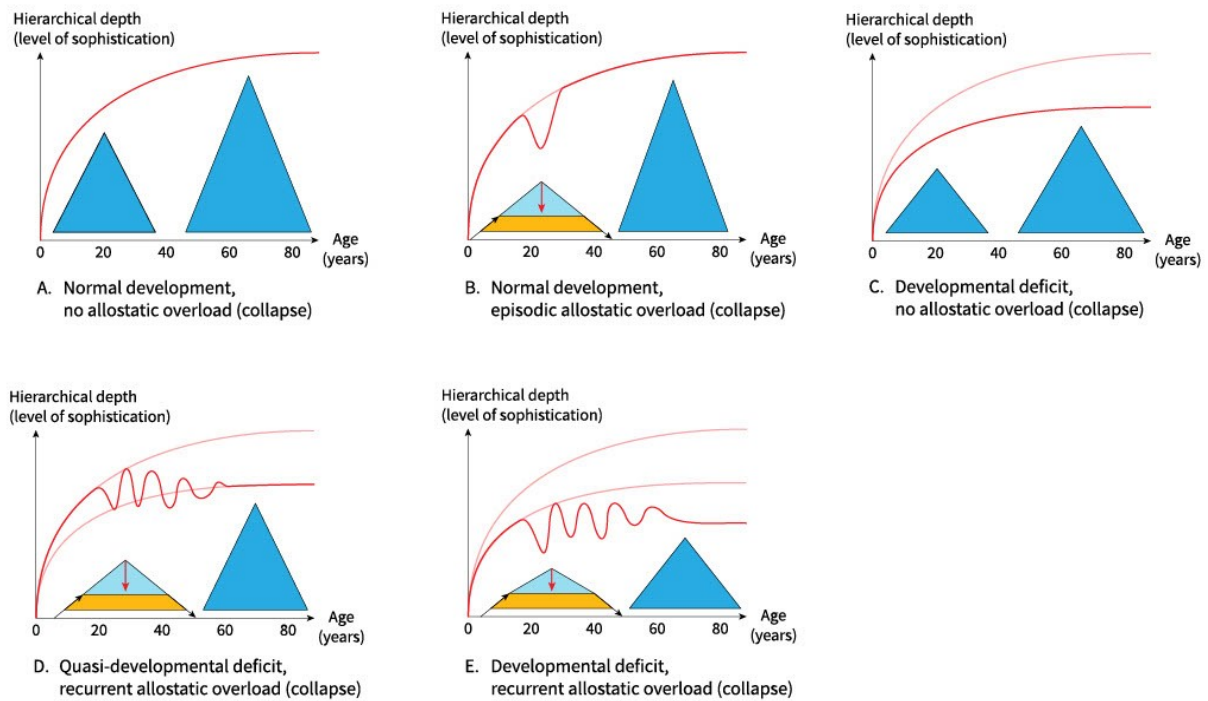

**Figure 9: Four logical deficits of hierarchical control in living systems across the lifespan**

*Curve shows hierarchical depth against time.*

- A. Optimal development, no episodic or chronic deficit
- B. An (episodic) loss of hierarchical control relative to a previous attained level
- C. A developmental (chronic) deficit of hierarchical control without episodic collapses.
- D. A prolonged sequence of episodic collapses, interfering with development
- E. A developmental deficit leading to a prolonged sequence of episodic collapses (vicious cycle)

In summary, we propose that the mereological (nested modular, hierarchical) structure of living system imposes a power-law imprint onto its dynamics (1/f 'pink noise'). This theory may have several consequences. Rather than a mere by-product, pink noise appears to be a cardinal feature of hierarchical message passing (information processing) in living systems. The current theory has implications for studies of the ontogeny (developmental aspects), phylogeny (evolutionary aspects) and the practical management of living systems in clinical medicine and ecology. These implications will be discussed below.

## 6. Implications of the theory: ontological and phylogenetic aspects

In our view, power-law frequency spectra reflect the hierarchical structure of a system: higher (mereological) levels exert a tonic pressure onto the dynamics of their (constituent) lower-level systems through cross-frequency coupling ('top-down control'). The top of a regulatory hierarchy is therefore responsible for producing the zero-frequency 'offsets' (DC-components) in the timeseries of a system's overt behavior (Figure 10). Such offsets reflect the default or mean expression levels of a system's inner message passing and overt behavior (Figure 10). In other words, the top of a regulatory hierarchy controls the behavioral 'climate' of an organism (its personality) as opposed to its behavioral 'weather' (fast changing action-perception sequences), which is produced at its base. This conclusion is supported by studies in different fields of science. For instance, translational studies show that nearly all species express inter-individual differences in behavioral traits, i.e. 'personalities' (148, 149).

The cause of such differences has been disproportionately localized within higher regulatory areas. For instance, high-level hub neurons and brain regions have been implicated in controlling stable behavioral traits in nematodes, smaller animals and humans (150-152). In bacteria, individuals specimens from the same strain may differ with respect to exploratory or social behavior, which has been linked to differential expression levels of higher-level regulatory genes in genetic information bottlenecks (153-158). In plants, regulatory genes have been implicated in controlling individual differences in plant communication strategies, causing scientists to consider the possibility of 'plant personalities' (i.e. more exploratory or avoidant growth patterns) (159). Finally, human studies show that hub brain regions controls the development of adult personalities and that developmental deficits or pathological changes in such areas produce personality-disorders (160, 161).

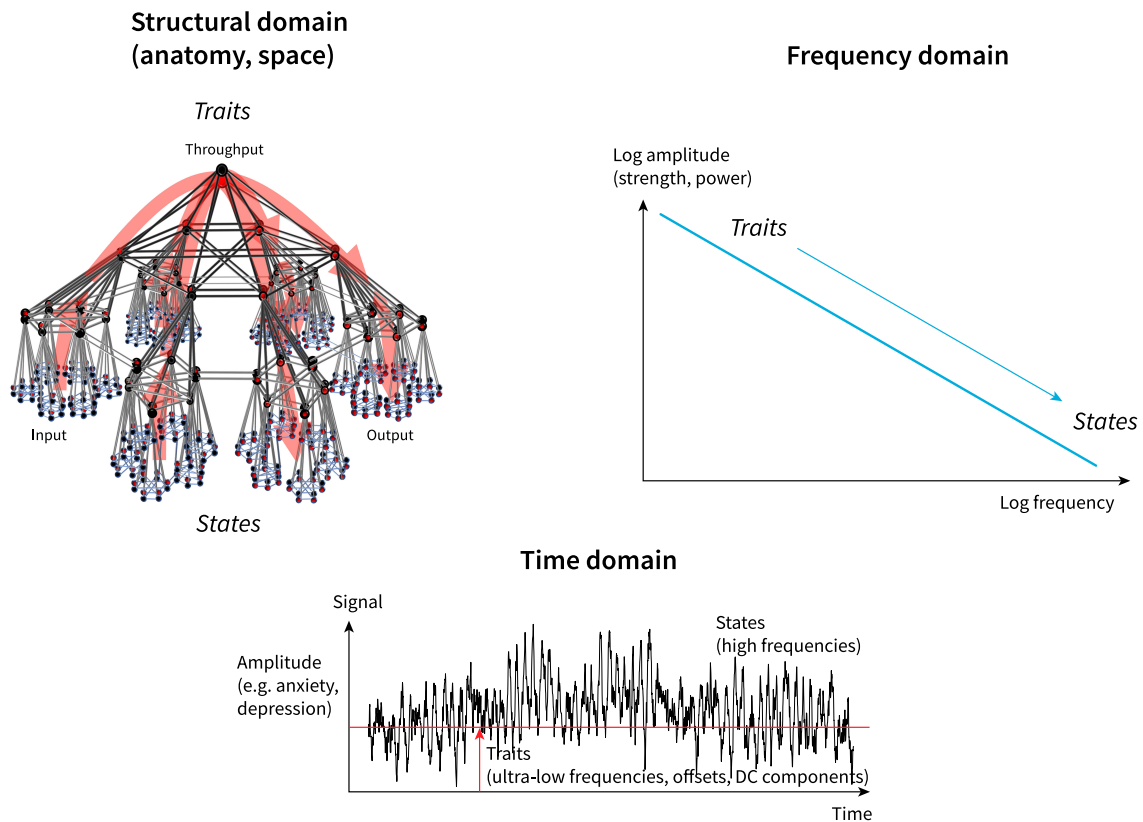

**Figure 10. How the top of a regulatory hierarchy controls the expression of stable aspects of inner experience and overt behavior (personality traits)**

*The top of a regulatory hierarchy contains high-level priors (setpoints, 'thermostats') that anticipate certain stable aspects of the environment (e.g. how safe or violent an econiche will be). High-level priors are rarely updated by low-level prediction errors, causing their values to be relatively stable. Thus, high-level priors impose ultra-low frequency signal components onto the timeseries of their subordinate systems, down to the level of simple action-perception cycles (reflexes), which are involved e.g. in producing fight, flight, or freezing responses in living systems (such stable signal components are called 'offsets' in statistics, or 'DC components' in engineering). Individual differences in the values of high-level priors produce more anxious (shy) or aggressive (dominant) individuals. Thus, high-level regulatory systems control the emergence of personalities in living systems.*

Apart from controlling overt behavior such as locomotion or fight- or flight responses, high-level regulatory systems control the physical aspects of organisms. For example, homeotic genes are regulatory genes that control the timing and duration of gene expression, protein synthesis, cell division, and cell migration during embryogenesis. This eventually governs the outgrowth of e.g. arms, legs, spines, tails or wings, producing different physical phenotypes (162). Other regulatory genes control the transition from the body plans of youngsters into the adult phenotype during ontogenesis. Individual differences in the expression of such genes explain within-specifies differences in physical appearance (163). Problems in regulatory genes involved in physical development (e.g. growth hormone receptor) may lead to physical malformations, depending on the type of gene or tissue that is involved (163, 164). Thus, high-level regulatory areas in many cases control within-species inter-individual differences in morphological traits (body plans) as well as behavioral traits (personalities), which is collectively referred to as 'phenotypic variance'.

As observed, high-level regulatory systems contain prior units that encode the abstract and temporally stable aspects of an organism's econiche (such as meteorological or social climates). In other words, phenotypic variance reflects an individual's tendency to anticipate physically or behaviorally to certain stable aspects of the environment, i.e. a degree of *specialization* (165). Such specialization allows for a 'division of labor' that prevents organisms from having to compete for similar (social) econiches (166). In most organisms, morphological traits covary with personality traits to produce an overall phenotype and corresponding division of labor (e.g. in ants, individual differences in the body size, strength and aggression ants causes them to take on different social roles such as workers, nurses and soldiers). Such phenotypes are controlled by a limited number of regulatory hub genes that form the knot of a bowtie motif (167, 168). Likewise, many species display clear sex differences in morphology and behavior (sexual dimorphism), which allows for a division of labor with respect to the production and rearing of offspring and fending off threats (169, 170). Such differences are driven by a differential expression of high-level regulatory hub genes during embryogenesis and ontogenesis (171, 172). In short, phenotypic variance of organisms reflects a degree of specialization that in many cases is controlled by high-level regulatory areas (information bottlenecks, the knots of bowties), rather than the base of the hierarchy (the wings of a bowtie).

As discussed, high-level regulatory areas are information bottlenecks that consist of a small number of hub nodes (a rich club) that control large areas of the network (5). Because of their central positions, only minor changes in such key areas suffice to produce widely different body plans and personalities. When such variants are actively selected upon, novel species may emerge. This idea is supported by a growing number of observations that link bowtie motifs to the robustness and evolvability of living systems (47). For instance, studies show that evolutionary changes in the past often involved the introduction of small mutations in so called 'hotspot genes' that are overrepresented in regulatory areas (172-174). Studies in cichlids and voles show that diverse species may quickly emerge from a common ancestor as a result of only minor modifications in high-level regulatory systems, leading to a wide variety of body plans, diet specializations, social preferences and corresponding geographical distributions: a phenomenon known as adaptive radiation (175-177). Similar findings involve the evolution of social behavior in different ant species (168, 178, 179) and plant species (180). In mammals, most between-species variance in brain architecture is found within mesocortical, prefrontal and anterior temporal areas, which involve regulatory (hub) regions. In contrast, sensorimotor cortices show relatively little cross-species differences (181). Rather than the base of the hierarchy (the wings of a bowtie), therefore, it's the *regulation* of such basic machinery (the knot of the bowtie) that makes the difference. A subtle tweaking of regulatory systems may be sufficient to distribute organisms across widely different econiches and social roles to optimize survival rates. Information bottlenecks therefore seem to play a key role in the specialization and speciation of living systems, which puts them center stage as 'hotspots of evolution' (Figure 11).

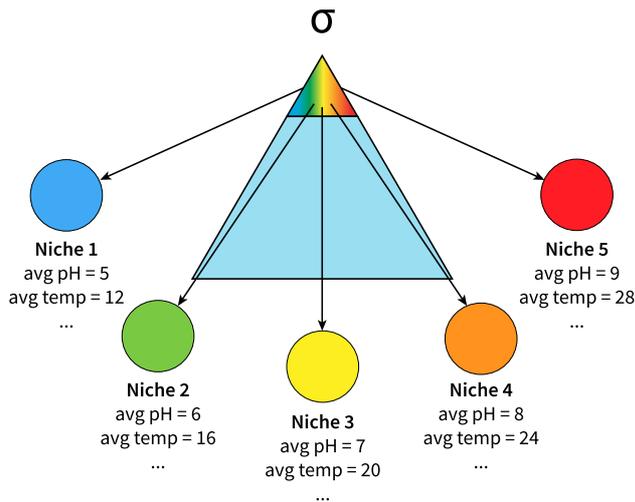

**Figure 11. Information bottlenecks as 'hotspots of evolution'**

*Subtle alterations at the top of a regulatory hierarchy (the knot of a bowtie) produce a wide spectrum of different phenotypes (physical traits and personality traits) that may occupy a range of different ecological niches. Such phenotypical spectra are subjected to natural selection to quickly produce new species (adaptive radiation).*

To summarize, we propose that power-law frequency distributions result from the hierarchical (mereological) architecture of living systems. The top of the hierarchy controls the expression of behavioral as well as physical traits (i.e., the stable phenotype). Subtle modifications to information bottlenecks may cause strong phenotypic variance and drive the specialization and speciation of living systems.

**7. Implications of the theory: practical aspects**

The idea that power-law frequency spectra convey information about the hierarchical dynamics of living systems has several practical consequences. For instance, such spectra could be used to monitor the quality of hierarchical information processing in living systems. In a previous study, we showed that the overburdening of regulatory capacity of living systems leads to a metabolic failure of hub nodes at the top of the regulatory hierarchy, producing a top-down collapse of hierarchical control ('allostatic overload' (34, 80)). This produces a form of desynchronization, leaving subordinate levels in an underregulated state of disorder. Such disorder can be quantified by a single measure called 'permutation entropy', which expresses the amount of randomness or unpredictability in the timeseries of a system. This measure captures the major hallmarks of the process of critical slowing down: a situation in which overburdened regulatory systems become slow to recover from perturbation (80). Increases in permutation entropy scores have been used as an early warning sign to discrete points that mark the transition from healthy and stable behavior to various physical and mental 'disorders', allowing for precautionary measures. Such tipping points (bifurcations or catastrophes) occur in any system under a significant amount of stress. Rising levels of permutation entropy have been used to predict the collapse of single organisms as well as social systems, ecosystems, earthquakes, landslides, and stock market crashes (80, 182). Although living and non-living systems differ with respect to their higher order statistics of power-law frequency distributions (7), we expect these phenomena to nonetheless connect at the level of energy dissipation, which powers the dynamics of all open systems, whether living or non-living (66, 120). Likewise, permutation entropy may be used as a universal quantifier of 'disorder' in any open dissipative system (80).

To permutation entropy, we now add a second measure from which to judge the quality of hierarchical information processing, which is the (steepness of the) power-law frequency distribution. In our view, a collapse of hierarchical control leads to a proportional loss of lower frequencies (and possibly an increase in higher frequencies), which is signaled by a sudden flattening of the power-law curve (Figure 8D). Like permutation entropy, this flattening may serve as an early warning sign for tipping points that mark the transition from ordered (healthy) states to disordered (unhealthy) states. These predictions can be tested in any hierarchical (Bayesian) control system that is taxed beyond its abilities to correct for environmental disturbances. Monitoring power-law exponents in relation to entropy (disorder) scores may thus serve preventive purposes (e.g. predicting and preventing heart attacks, arrhythmias, epileptic seizures or mental disorders, relapses in autoinflammatory diseases, patients with mental disorders, social systems in disarray, polluted ecosystems, but also failing governments, earthquakes, and stock market crashes) or to seek out weak spots in a system's ability to control certain environmental challenges (e.g. selecting suitable antibiotics to treat bacterial infections (183)).

Power-law dynamics may also serve to monitor the effects of interventions into dynamic systems. The degree to which power-law curves recover, disorder levels diminish and symptoms abate after certain interventions may serve as a quantifier for the restoration of hierarchical functioning and therapeutic success in clinical medicine. In psychiatry, for instance, the balance between top down and bottom up information processing (hierarchical depth) may be tuned by neuromodulatory neurotransmitter systems, using pharmacotherapeutic agents such as antidepressants or other psychoactive drugs that modulate the efficiency of neural processing. The effect of such interventions can be tested by examining changes in power-law curves of brain function before and after treatment.

## 8. Discussion

We have proposed that hierarchical structure may produce hierarchical dynamics (state-trait continua) in (open dissipative) systems of coupled oscillators. Such systems engage in a vertical encoding of the deep spatiotemporal structure of their environments, which explains the emergence of different frequencies: strongly coupled (hub) nodes at the top of a hierarchy produce low frequencies whereas loosely coupled (peripheral) nodes at its base produce high frequencies. The typical run-off of amplitude with frequency is explained by an inverse relationship between coupling strength and energy dissipation rate. Crucially, this proposal applies both to living and non-living systems. In the resting (not actively exploring) state, living systems display smooth power-law curves that represent an equal dissipation of energy across all hierarchical levels and corresponding frequency bands. When actively sampling their environments (during task performance), living systems develop bumps on their power-law curves. In our view, such bumps reflect the amplification of frequencies produced at a self-organized optimal hierarchical depth of information processing. Extreme task performance selectively overburdens the top of a regulatory hierarchy, which is due to the overrepresentation of strongly dissipating (metabolically demanding) hub nodes in such areas. When such nodes overload and fail, this causes a top-down cascading failure of central-integrative hierarchical control ('allostatic overload'). This in turn causes a loss of synchronization (disorder) of signal changes at lower levels, which can be quantified in terms of the permutation entropy score of the timeseries of a system. Such changes should coincide with a flattening of low-normalized powerlaw curves, reflecting decreased involvement of higher levels of control (lower frequencies). In this view, increased permutation entropy describes the more distal effects of collapsing hierarchical control (i.e. disorder), whereas flattening slopes may convey more proximal information (i.e. dysregulation: collapse of hierarchical control). Together, such changes may serve as early warning signs to system failure.

Testing this theory requires a transdisciplinary approach, which should avoid subjecting living creatures to unnecessarily stressful circumstances. One can consider studies in microbes under antibiotic challenges, plants under conditions of severe drought, or situations in which 'natural experiments' have brought living systems to the limits of their regulatory capacities, such as patients suffering from various physical or mental conditions. An open question is the degree to which the proposed measures of the sophistication of hierarchical control apply to large scale social systems such as ant colonies, beehives, human communities, biotopes, organizations and governments. All such systems may display low-frequency traits or 'personalities', i.e. a default expression-level of certain policies that reflect certain stable aspects of the environment (e.g. country, culture) and, hence, some degree of specialization. Studies that address such questions can make use of *in silico* models to simulate the effects of changing hierarchical structures on the dynamics of a system. Such studies require models of deep oscillatory neural networks with a folded information bottleneck motif that are involved in Bayesian inference. To date, many of such models are yet to be developed (for an early exception, see (184)). Because of its scale free nature, the free energy principle and its corollary, active inference, have the exciting prospect marrying the physical sciences and the humanities, with obvious philosophical and ethical implications (185). Overall, the 21st century promises to be an extraordinary time for anyone who wishes to optimize their predictive models of the world.

**Acknowledgements**

We thank the reviewers for their constructive comments.